\documentclass[12pt,reqno]{article}
\usepackage{amsmath,epsfig}

\allowdisplaybreaks
\numberwithin{equation}{section}

\DeclareSymbolFont{AMSa}{U}{msa}{m}{n}
\DeclareSymbolFont{AMSb}{U}{msb}{m}{n}
\DeclareMathSymbol{\fieldR}{\mathalpha}{AMSb}{"52}

\begin{document} 

\begin{flushright} \small
hep-th/0412183 \\ KCL-MTH-04/16 \\ ITP--UU--04/51 \\ SPIN--04/33 \\ UG-04/04
\end{flushright}
\bigskip

\begin{center}
 {\large\bfseries Non-extremal instantons and wormholes in string theory} 
\\[5mm]
 E. Bergshoeff$^1$, A. Collinucci$^1$, U. Gran$^2$, D. Roest$^2$ and 
S. Vandoren$^3$ \\[3mm]
  {\small\slshape
 $^1$ Centre for Theoretical Physics, University of Groningen, Nijenborgh 4, 
9747 AG Groningen, The Netherlands \\
 {\upshape\ttfamily e.a.bergshoeff, a.collinucci@phys.rug.nl}\\[3mm]
 $^2$ Department of Mathematics, King's College London, Strand,
London, WC2R 2LS, United Kingdom \\
 {\upshape\ttfamily ugran,droest@mth.kcl.ac.uk}\\[3mm]
 $^3$ Institute for Theoretical Physics \emph{and} Spinoza Institute \\
 Utrecht University, 3508 TD Utrecht, The Netherlands \\
 {\upshape\ttfamily s.vandoren@phys.uu.nl}}
\end{center}
\vspace{5mm}

\hrule\bigskip

\centerline{\bfseries Abstract} \medskip

We construct the most general non-extremal spherically symmetric 
instanton solution of a gravity-dilaton-axion system with $SL(2,R)$ symmetry,
for arbitrary euclidean spacetime dimension $D\geq 3$. A subclass of these 
solutions
describe completely regular wormhole geometries, whose size is 
determined by an invariant combination of the $SL(2,R)$ charges. 

Our results can be applied to four-dimensional effective actions of type II 
strings compactified on a Calabi-Yau manifold, and in particular to the 
universal 
hypermultiplet coupled to gravity. We show that these models contain 
regular wormhole solutions, supported by regular 
dilaton and RR scalar fields of the universal hypermultiplet. 

\bigskip

\hrule\bigskip

\section{The action}

This paper contains a summary of \cite{Bergshoeff:2004fq}. In addition, we
demonstrate the existence of regular wormhole solutions in the universal
hypermultiplet, which is present in the low-energy effective action 
of type II string theories compactified on a Calabi-Yau manifold. Wormhole 
solutions from Calabi-Yau compactifications 
were also found in \cite{Giddings:1989bq}. The main difference 
with our solution is that in our case we also include RR fields, and 
furthermore, our solution is regular in the complete domain of the wormhole 
geometry.

We start with the action of a gravity-dilaton-axion system in 
$D$-spacetime dimensions. In Minkowski space, the Lagrangian is
\begin{equation}
  \mathcal{L}_M  = \tfrac{1}{2} \sqrt{|g|} \,
    [ {R} -\tfrac{1}{2} (\partial {\phi})^2
    -\tfrac{1}{2} e^{b {\phi}} (\partial {\chi})^2 ] \, ,
\label{10DIIB}
\end{equation}
where $b$ is an arbitrary dilaton coupling parameter. This Lagrangian has
an $SL(2,R)$ group of symmetries. They can be realized as modular 
transformations on the complex field
\begin{equation}
\tau \equiv \frac{b}{2}\, \chi + i\, e^{-b\phi/2}\ ,\qquad
\tau \rightarrow \frac{\alpha \tau +\beta}{\gamma\tau +\delta}\ ,\qquad  
\alpha\delta-\beta\gamma=1\ ,\label{tau}
\end{equation}
and is valid for any nonzero value of the dilaton coupling $b$.

This theory occurs for example as the scalar section of IIB supergravity in 
$D=10$ Minkowski space-time with dilaton-coupling parameter $b=2$. Other 
values of $b$ can arise when considering (truncations of) compactifications of 
II supergravity. The main example we will discuss here is that of the 
universal hypermultiplet, that arises after compactifying type IIA strings on 
a (rigid) Calabi-Yau threefold down to $D=4$. This hypermultiplet contains 
four scalars, $\phi$ and $\sigma$ coming from the NS sector, and $\psi$ and 
$\varphi$ coming from the RR sector. The four-dimensional Lagrangian can be 
written as
\begin{equation}\label{UHM}
\mathcal{L}_M  = \tfrac{1}{2} \sqrt{|g|} \,
    [ {R} -\tfrac{1}{2} (\partial {\phi})^2
    -\tfrac{1}{2} e^{\phi} \big((\partial {\psi})^2+(\partial {\varphi})^2\big)
-\tfrac{1}{2} e^{2\phi}\big(\partial \sigma +\psi \partial \varphi\big)^2] \, .
\end{equation}
The scalar symmetry group is now $SU(2,1)$, but contains various inequivalent 
$SL(2,R)$ subgroups. For instance, if we set both $\sigma=\psi=0$, we get 
(\ref{10DIIB}) whith $b=1$ whereby $\varphi$ is identified with $\chi$. 
If we set $\psi=\varphi=0$, we have $b=2$ and $\sigma$ is identified with 
$\chi$. By writing down the field equations for (\ref{UHM}), it 
is easy to see that these truncations are consistent.
Extremal instantons in the universal hypermultiplet have been discussed in 
detail in \cite{Theis:2002er,Davidse:2003ww,Davidse:2004gg}, and correspond 
to wrapped 
(euclidean) membranes along three-cycles, or wrapped NS5-branes along the 
entire Calabi-Yau. These two cases correspond 
to $b=1$ and $b=2$ respectively. Using the results obtained in 
\cite{Bergshoeff:2004fq}, we will here generalize this to the 
non-extremal case, and show that there are interesting and new solutions
that have the spacetime geometry of a wormhole.

To discuss instantons, we first have to perform a Wick rotation. This rotation
is best understood by dualizing the axion into a $(D-2)$-form potential. 
One then finds that under a Wick rotation, $\chi \rightarrow i \chi $.
The Euclidean Lagrangian corresponding to  (\ref{10DIIB}) is then
\begin{equation}
\mathcal{L}_E
  = \tfrac{1}{2} \sqrt{g} \,
    [ {R} -\tfrac{1}{2} (\partial {\phi})^2
    +\tfrac{1}{2} e^{b {\phi}} (\partial {\chi})^2 ] \, ,
\label{EuclideanAction}
\end{equation}
with all fields real. Notice that in the scalar formulation, as opposed to
the formulation with the $(D-1)$-form field strength, the contribution to the 
action coming from the scalar sector is not positive definite.
For $b=2$ and $D=10$ this is the gravity-scalar part of 
Euclidean IIB supergravity, in which the D-instanton can easily be found
as a solution of the Euclidean equations of motion 
\cite{Gibbons:1996vg,Green:1997tv}. The non-extremal solutions were
found in \cite{Bergshoeff:2004fq}, and we repeat them in the next section.

As already explained, compactifications of string theory can give rise to 
other values 
of $b$. The Euclidean version of the universal hypermultiplet Lagrangian
(\ref{UHM}) can best be understood in terms of the double-tensor multiplet 
formulation, in which $\varphi$ and $\sigma$ are dualized 
into two antisymmetric tensors \cite{Theis:2002er}. After a Wick rotation, 
$\varphi \rightarrow i\varphi, \sigma \rightarrow i\sigma$, and the 
Euclidean Lagrangian for the universal hypermultiplet becomes
\begin{equation}\label{EUHM}
\mathcal{L}_E  = \tfrac{1}{2} \sqrt{|g|} \,
    [ {R} -\tfrac{1}{2} (\partial {\phi})^2
    -\tfrac{1}{2} e^{\phi} \big((\partial {\psi})^2-(\partial {\varphi})^2\big)
+\tfrac{1}{2} e^{2\phi}\big(\partial \sigma +\psi \partial \varphi\big)^2] \, .
\end{equation}
Notice that the two truncations, $\psi = \sigma =0$ and $\psi=\varphi=0$,
both fall into the class of (\ref{EuclideanAction}), in which we have
$b=1$ and $b=2$ respectively.

There are three conserved currents for the $SL(2,R)$ transformations in the 
Euclidean model, satisfying  $\nabla_\mu j^\mu=0$.
The corresponding charges are denoted by $q_3,q_+$ and $q_-$, and are 
normalized as specified in \cite{Bergshoeff:2004fq}. They transform under 
$SL(2,R)$ in such
a way that the combination
\begin{equation}
{\bf \mathsf{q}}^2 \equiv q_3^2 - q_+q_-\ ,
\end{equation}
is invariant \cite{Bergshoeff:2002mb,Bergshoeff:2004fq}. 
The three conjugacy classes of 
$SL(2,R)$ then correspond to ${\bf \mathsf{q}}^2 < 0, {\bf \mathsf{q}}^2=0$ and ${\bf \mathsf{q}}^2 > 0$. 
The extremal solutions will have ${\bf \mathsf{q}}^2=0$, the non-extremal ${\bf \mathsf{q}}^2 
\neq 0$. The wormhole solutions will have ${\bf \mathsf{q}}^2 < 0$.
For later convenience, it is useful to define the quantity
\begin{equation}
c\equiv {\sqrt {\frac{2(D-1)}{D-2}}}\ ,
\end{equation}
which will appear explicitly in the instanton solutions below.

\section{Instanton solutions}

We search for generalised instanton solutions with manifest
$SO(D)$ symmetry,
\begin{align}
  {ds}^2 & = e^{2\,B(r)} (dr^2 + r^2 d\Omega_{D-1}^2) \,, \qquad
\phi=\phi(r) \,, \qquad \chi=\chi(r)\,.
  \label{instanton}
\end{align}
The standard D-instanton solution \cite{Gibbons:1996vg} is
obtained for the special case that $B(r)$ is constant. Other references
on generalised instantons and wormholes that are related
to our work are \cite{Giddings:1989bq,Rey:1989xj,Coule:1989xu,Kim:1997hq,
Bergshoeff:1998ry,Einhorn:2002am,Gutperle:2002km,Einhorn:2002sj,Kim:2003js,
Maldacena:2004rf}.
To obtain an $SO(D)$ symmetric generalised instanton solution, we
allow for a non-constant $B(r)$ and solve the field equations
following from the Euclidean action (\ref{EuclideanAction}). This
was done in detail in \cite{Bergshoeff:2004fq}. Here we summarise the result.

The solution can be written in a compact
form by using a harmonic function $H(r)$ over a conformally flat space with 
metric as given in (\ref{instanton}),
\begin{align}
  H(r)= \frac{b\,c}{2}\,\log(f_{+}(r)/f_{-}(r))\,, \quad B(r)=\frac{1}{D-2}
\log(f_{+}f_{-})\,,\quad
f_{\pm}(r) = 1\pm\frac{\mathsf{q}}{r^{D-2}}\,,
\end{align}
The general instanton solution can then be written as 
\begin{equation}\label{SolEq}
\boxed{
\begin{aligned}
  ds^2 & = \left(1-\frac{\mathsf{q}^2}{r^{2\,(D-2)}}\right)^{2/(D-2)}\,(dr^2
+ r^2 d
  \Omega_{D-1}^2) \,, \\
  e^{b\,\phi(r)} & =
\left(\frac{q_{-}}{\mathsf{q}}\,\sinh(H(r)+C_1)\right)^2\,,\\
  \chi(r) & =
  \frac{2}{b\,q_{-}}\,(\mathsf{q}\,\coth(H(r)+C_1)-q_3)\,.
\end{aligned}
}
\end{equation}
This solution is valid for any value~\footnote{The case 
$b=0$ is treated in \cite{Myers:1988sp}.} of $b\neq 0$.
The integration 
constant $C_1$ can be traded for the asymptotic value of the dilaton
that we will later identify with the string coupling constant. 
Notice also the explicit dependence on the $Sl(2,R)$ charges $q_3,q_-$ and 
$q_+$. The solutions \eqref{SolEq} are valid both for $\mathsf{q}^2\equiv 
q_3^2-q_-q_+$ positive, negative and zero, corresponding to the three
conjugacy classes of $SL(2,R)$. We now discuss these three cases separately:

\begin{itemize}
\item
 $\bf \mathsf{q}^2 >0$: Black Holes

In this case $\mathsf{q}$ is real and the solution is given by
\eqref{SolEq} with all constants real.
However, the metric becomes imaginary below a critical radius
 \begin{equation} \label{rcritical}
  r^{D-2} < r_c^{D-2} = \mathsf{q} \, .
 \end{equation}
One can check that there is a curvature singularity at $r=r_c$, which 
happens at strong string coupling: $e^{\phi(r)} \rightarrow \infty$
as $r \rightarrow r_c$.

Between $r=r_c$ and $r=\infty$, $H$ varies between $\infty$ and $0$, and with
an appropriate choice of $C_1$, i.e. a positive value of $C_1$, the scalars 
have no further singularities in this domain.
Thus one might hope to have a modification of this solution by higher-order 
contributions to the effective action of IIB string theory 
\cite{Einhorn:2002am}. Alternatively, one can consider the possible resolution 
of this singularity upon uplifting to one higher dimension. 
In \cite{Bergshoeff:2004fq}, we showed that this indeed happens for the 
special case of
\begin{equation}
b \geq \sqrt{\frac{2(D-2)}{D-1}} \,,
\end{equation}
equivalent to $bc \geq 2$. Upon uplifting, this becomes a non-extremal 
dilatonic black hole. The case when $bc=2$ lifts up to a (non-dilatonic)
Reissner-Nordstr\"om black hole with mass and charge given by 
\begin{align}
Q&= -2\,q_{-} \,, \qquad M=2\,\sqrt{\mathsf{q}^2+q_{-}^2}\qquad
\Rightarrow \qquad \mathsf{q}^2 = \frac{M^2-Q^2}{4}\, .
\label{RN-instanton-relation}
\end{align}
Hence, the ${\bf \mathsf{q}}^2>0$ solutions with $bc\geq 2$ are spatial sections of a 
higher-dimensional (Lorentzian) black hole solution. The case of $bc < 2$
cannot be uplifted and remain singular instanton solutions in $D$-dimensions.
In Einstein frame, these geometries are singular wormholes that are pinched at
the selfdual radius $r_{{\rm sd}}=r_c$ \cite{Bergshoeff:2004fq}.

In the case of $\mathsf{q}^2 > 0$, there is an interesting limit
in which $q_- \rightarrow 0$. This yields a 
solution with only two independent
integration constants, $q_+$ and $\mathsf{q}^2$. The range of
validity of this solution is equal to that of the above solution
with $q_- \neq 0$: it is well-defined for $r>r_c$, while at $r =
r_c$ the metric has a singularity and the dilaton blows up. The
singularity can be resolved upon uplifting for all
values of $bc \geq 2$ to Schwarzschild black holes, with mass 
$M=2{\bf \mathsf{q}}$. More details can be found in \cite{Bergshoeff:2004fq}.

\item $\bf \mathsf{q}^2 =0$: Extremal instantons

 We now consider the limit $\mathsf{q}^2 \rightarrow 0$ of
the general solution \eqref{SolEq}, after rescaling the constant $C_1$
with a factor $\mathsf{q}$ to make the limit well-defined. 
Taking the limit yields the extremal solution:
\begin{equation}
\boxed{
\begin{aligned}
  ds^2 = dr^2+r^2\,d\Omega_{D-1}^2 \,, \qquad
  e^{b\,\phi(r)/2} = h \qquad \chi(r) = \frac{2}{b}\,(h^{-1} -
\frac{q_3}{q_-}) \,,   \label{instlimEq}
\end{aligned}
}
\end{equation}
where $h(r)$ is the harmonic function:
\begin{align}
  h(r) = g_s^{b/2} +  \frac{b\,c\,q_-}{r^{D-2}} \, ,
\end{align}
and $g_s$ is the asymptotic value of the dilaton at infinity.

This is the extremal D-instanton solution of \cite{Gibbons:1996vg}. This solution is
regular over the range $0 < r < \infty$ provided one takes both $g_s$ and $b\,c\,q_-$
positive; at $r=0$ however, the harmonic function blows up and the scalars are singular. Similar to the case of $\bf \mathsf{q}^2 >0$, these singular solutions
can be lifted to higher dimensions where, e.g. for $bc=2$, they become
extremal Reissner-Nordstr\"om black holes.

\item $\bf \mathsf{q}^2 <0$: Wormholes

In this case $\mathsf{q}$ is imaginary. To obtain a real solution
we must take $C_1$ to be imaginary. We therefore redefine
\begin{equation}
\mathsf{q}\rightarrow i\,\mathsf{\tilde{q}} \hskip 2truecm C_1
\rightarrow i\,\tilde{C_1}\, ,
\end{equation}
such that $\mathsf{\tilde{q}}$ and $\tilde{C_1}$ are real. One can
now rewrite the solution \eqref{SolEq} by using the
relation
 \begin{equation}
  \log(f_{+}/f_{-}) = 2 \,{\rm arctanh}(\mathsf{q}/r^{D-2})\, ,
 \end{equation}
and, next, replacing the hyperbolic trigonometric functions by
trigonometric ones in such a way that no imaginary quantities
appear. We thus find that, for $\mathsf{q}^2 <0$, the general
solution \eqref{SolEq} takes the following form:
\begin{equation}
\boxed{
\begin{aligned}
  ds^2 & =
(1+\frac{\mathsf{\tilde{q}}^2}{r^{2\,(D-2)}})^{2/(D-2)}\,(dr^2+r^2\,d\Omega_
{D-1}^2)\,,\\
  e^{b \phi(r)} & = \left(\frac{q_{-}}{\mathsf{\tilde{q}}}\,
  \sin(b\,c\,\arctan(\frac{\mathsf{\tilde{q}
}}{r^{D-2}})+\tilde{C_1}) \right)^2 \,,\\
  \chi(r) & =\frac{2}{b\,q_{-}}\,(\mathsf{\tilde{q}}\,
  \cot(b\,c\,\arctan(\frac{\mathsf{\tilde{q}}}
{r^{D-2}})+\tilde{C_1})-q_3)\,.
 \label{qminussol}
\end{aligned}
}
\end{equation}
The metric and curvature are well behaved over the range $0<r<\infty$. However, the
scalars can only be non-singular over the same range by an appropriate choice of
$\tilde{C}_1$ provided that $bc < 2$. This can be seen as follows. The $\arctan$ varies
over a range of $\pi/2$ when $r$ goes from $0$ to $\infty$. It is multiplied by $bc$ and
thus the argument of the sine varies over a range of more than $\pi$ if $bc > 2$.
Therefore, for $bc>2$ there is always a point $r_c$ such that $\chi \rightarrow \infty$
as $r \rightarrow r_c$. Note that the breakdown of the solution occurs at weak string
coupling: $e^{\phi} \rightarrow 0$ as $r \rightarrow r_c$. 
This singularity is not resolved upon uplifting and corresponds to a black hole
with a naked singularity (in the case of Reissner-Nordstr\"om, $M^2<Q^2$. 
The same holds for the liming case of $bc=2$. Therefore the case
$\mathsf{q}^2 <0$ only yields regular instanton solutions for $bc < 2$, together with the
condition that ${\tilde C}_1$ and ${\tilde C}_1+bc\pi/2$ are on the same branch of the cotangent.

The metric in (\ref{qminussol}) has a $Z_2$ isometry corresponding 
to the reflection $r^{D-2}\rightarrow \mathsf{\tilde{q}} r^{2-D}$
which interchanges the two asymptotically flat regions. This reflection
has a fixed point, corresponding to the selfdual radius
 \begin{align}
  r_{\text{sd}}^{D-2}=\mathsf{\tilde{q}} \,.
\end{align}
Furthermore, the 
thickness of the neck was in \cite{Bergshoeff:2004fq} computed to be
\begin{align}
 \rho_{\text{sd}}^{D-2} = 2\mathsf{\tilde{q}} \,.
 \end{align}
We have summarised this in the following figure:

\begin{figure}[ht]
\centerline{\epsfig{file=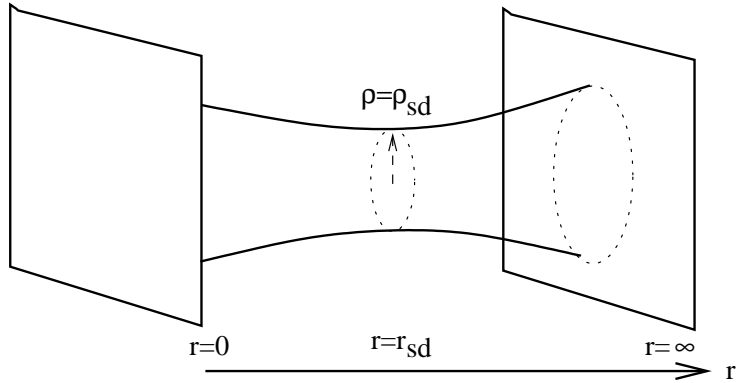,width=.6\textwidth}}
 \caption{\it The geometry of a wormhole. The two asymptotically flat
regions at $r=0$ and $r=\infty$ are
 connected via a neck with a minimal physical radius $\rho_{\text{sd}}$ at
the self-dual radius $r_{\text{sd}}$.}
 \label{fig:wormhole}
\end{figure}

\end{itemize}

\section{Instanton action}

The value of the action, evaluated on the instanton solution, is a key 
ingredient in the semiclassical approximation of the euclidean path integral.
In \cite{Bergshoeff:2004fq} we computed the instanton action for the three
cases, corresponding to $\mathsf{q}^2 >0, \mathsf{q}^2 =0$,
and $\mathsf{q}^2 <0$. This was done by specifying the additional 
surface term added to (\ref{EuclideanAction}), which solely determines
the instanton action. This surface term 
can be found from the 
dual description in terms of the $(D-1)$-form field strength formulation.
We here summarise the results.

For the case when $\mathsf{q}^2 \geq 0$, the contribution
to the action coming from infinity is given by
\begin{align} \label{nonextr-inst-act-infty}
 \mathcal{S}^{\infty}_{inst} = \frac{4}{b^2}\,(D-2)\, 
\mathcal{V}ol(S^{D-1})\,b\,c\,\Big(\sqrt{\mathsf{q}^2+\frac{q_-^2}{g_s^b}} 
\Big)\ .
\end{align}
Here we have used the relation between $C_1$ and the asymptotic value of
the dilaton, $g_s^{b}=(q_-/\mathsf{q})^2\,\sinh^2 C_1$. Notice that the 
instanton action is proportional to the mass of the black hole to which
the solution uplifts in one dimension higher. Furthermore, the 
result (\ref{nonextr-inst-act-infty}) also hols for $\mathsf{q}^2=0$,
which gives the lowest value of the action. The resulting instanton 
action is then inversely proportional to $g_s^{b/2}$. The D-instanton
of ten-dimensional IIB corresponds to taking $b=2$. The extremal 
instantons for the universal hypermultiplet action (\ref{UHM}) 
\cite{Theis:2002er,Davidse:2003ww}
also fall into this class: the membrane instantons correspond to $b=1$
whereas the NS-fivebrane instantons correspond~\footnote{To 
compare with \cite{Theis:2002er,Davidse:2003ww}, one has to redefine the 
string coupling constant by taking the square root. We correct
here a minor mistake in \cite{Bergshoeff:2004fq}, in which the membrane
and fivebrane instantons were written to correspond to $b=2$ and $b=4$.}
to $b=2$. We have here given only the contribution from infinitiy. 
The non-extremal instantons also contribute to the action at the other
boundary, where $r=r_c$. Since the solution is singular at this point, it is 
however not clear that the supergravity approximation is still valid in this 
region.

The case when $\mathsf{q}^2 < 0$ is very different. For $bc<2$, these
are regular wormhole solutions with
two asymptotic 
boundaries at $r=0$ and $r=\infty$ that are related by a reflection symmetry. 
The wormhole action gets contributions from both these boundaries, and 
the result is
\begin{align} \label{tildeq-inst-act}
 \mathcal{S}_{wormhole} = \frac{4}{b^2}\,(D-2) \mathcal{V}ol(S^{D-1})\,{b\,c}\,
\mathsf{\tilde {q}}\,\Big(\cot \tilde{C}_1 
-\cot (\tilde{C}_1 + bc \frac{\pi}{2})\Big)\,.
\end{align}
Due to the fact that $\tilde{C}_1$ and $\tilde{C}_1 + bc\pi/2$ are on the same 
branch of the cotangent, the total instanton action is manifestly positive
definite. One can rewrite the above result in terms of the string coupling
constant, using $g_s^{b/2}\equiv e^{b\phi_{\infty}/2}=(q_-/\mathsf{\tilde {q}})
\sin {\tilde C_1}$. In the neighborhood of $bc\approx 2$, the instanton 
action becomes very large, and in the limit to the critical point where 
$bc=2$, it diverges. At that point, the wormhole solution is no longer regular.

\section{Wormholes in string theory} 

We have seen that the condition for regular wormholes is that there exist
models for which $bc<2$. In type IIB in ten dimensions, this is not satisfied.
Toroidal compactifications of string theory only lead to values
of $b$ for which $bc\geq 2$, so no wormholes exist for these cases. 
However, we have seen that for the universal hypermultplet, which descends
from a Calabi-Yau compactification of type II strings, one can have the value $b=1$ in 
$D=4$, and so $bc={\sqrt 3}<2$. The solution is then characterized by the 
dilaton and the RR scalar $\varphi$ that descends from the RR three-form 
gauge potential in IIA in ten dimensions.
Since the extremal case $\mathsf{q}^2 =0$ corresponds to a wrapped type 
IIA euclidean membrane over a (supersymmetric) three-cycle, it is natural to 
suggest that the wormhole, with $\mathsf{q}^2 < 0$, corresponds to a wrapped 
non-extremal euclidean D2 brane. \\[3mm]

{\bf Acknowledgement}

It is a pleasure to thank Mathijs de Vroome for stimulating discussions.

This work is supported in part by the Spanish grant BFM2003-01090 and the 
European Community's Human Potential Programme under contract 
HPRN-CT-2000-00131 Quantum Spacetime, in which E.B. and D.R.~are associated 
to Utrecht University. The work of U.G.~is funded by the Swedish Research 
Council.

\end{document}